\newcommand{\be}{\begin{equation}}
\newcommand{\ee}{\end{equation}}
\newcommand{\bea}{\begin{eqnarray}}
\newcommand{\eea}{\end{eqnarray}}
\newcommand{\slp}{\slash{\!\!\!p}}
\newcommand{\slk}{\slash{\!\!\!k}}
\newcommand{\bm}[1]{{\mbox{\boldmath $ #1 $}}}

\documentstyle[preprint,aps,epsf]{revtex}
\tighten
\begin{document}
\draft 
\title{Complete spin structure of the pion-nucleon-loop delta 
self-energy
}
\author{C.~L.~Korpa}
\address{Department of Theoretical Physics, Janus Pannonius
University,
Ifj\'us\'ag u.\ 6, 7624 P\'ecs, Hungary}

\maketitle

\begin{abstract}
The complete spin structure of the pion-nucleon-loop contribution
to the delta self-energy and dressed
propagator is calculated in vacuum, with the most general form
of the pion-nucleon-delta vertex. The imaginary parts of the ten
Lorentz-scalar coefficients are calculated in closed form, while
the real parts are obtained numerically from a dispersion relation.
The effect of the pion-nucleon-delta coupling constants and 
form-factor on the pion-nucleon phase-shift in the spin-3/2
isospin-3/2 channel is studied.
\end{abstract}

%%%%%%%%%%%%%%%%%%%%%%%%%%%%%%%%%%%%%%%%%%%%%%%%%%%%%%%%%%%%%%%%%%%%% 
\section{Introduction}
%%%%%%%%%%%%%%%%%%%%%%%%%%%%%%%%%%%%%%%%%%%%%%%%%%%%%%%%%%%%%%%%%%%%%
The delta resonance plays a significant role in a number
of processes due to the fact that it is easily excited from a
nucleon, 
especially by pions.
The spin-3/2 value of the delta is responsible for
its somewhat cumbersome treatment in a fully relativistic
approach. As a consequence, either a nonrelativistic approximation
is used \cite{Brown}, or the (dressed) delta propagator is
approximated
by using a definite scheme for calculating its self-energy 
\cite{Rudi,Phil,Olaf}.

There is a general ambiguity in the Rarita-Schwinger form of 
the propagator of a 
spin-3/2 particle \cite{Nath}, which allows one to take for the
free propagator the usual form:
\be
G_0^{\mu\nu}(p)=\frac{\slp +M_\Delta}{p^2-M^2_\Delta+
i\varepsilon}
\left[ g^{\mu\nu}-\frac{\gamma^\mu\gamma^\nu}{3}
-\frac{2p^\mu p^\nu}{3M^2_\Delta}
+\frac{p^\mu\gamma^\nu-p^\nu\gamma^\mu}{3M_\Delta}
\right],\label{freepropagator}
\ee
where $M_\Delta$ is the (bare) mass of the delta. 
The most general form of the pion-nucleon-delta coupling is 
described by the interaction Lagrangian density
\be
{\cal{L}}_{\pi N\Delta}=
\frac{g}{M} \partial_\alpha \pi \bar{\Delta}_\beta
(g^{\alpha\beta}+a \gamma^\beta \gamma^\alpha)N + \rm{h.c.},
\label{pnd}
\ee
where $g$ is the (dimensionless) coupling, $M$ the physical mass of
nucleon (introduced in the coupling to make explicit its dimension),
$a$ an arbitrary (the so called off-shell) 
parameter and the isospin
coefficients and indices are not written out. In some treatments the
value of the parameter $a$ is taken to be zero for reasons of
simplicity, but we do not want to make such an {\it ad hoc\/} 
restriction.

The large value of the coupling $g$ means that the properties of
the nucleon, pion and delta are mutually strongly influenced. 
For the pion and the nucleon in vacuum the modification of
the spectral function occurs at rather large momenta, where 
finite-size (i.e. form-factor) effects are expected to 
suppress the correction. For the delta, however, the
pion-nucleon-loop
correction determines the vacuum spectral function in a decisive way,
as one can see from the delta decay properties.
Such a dressed delta
plays an important role in a number of processes with nucleon (or
nuclear) targets, motivating a fully relativistic calculation
of its self-energy and dressed propagator. The pion-nucleon loop is
expected to give the dominant contribution at low and intermediate 
energies, which is
the region most readily probed by experiments. 

It is probably even of greater importance to study the properties of 
the delta (and nucleon and pion) in the hot nuclear medium, as
created
for example in collisions of heavy ions, than in the relatively
simple
case of free space. The in-medium calculations at
present usually involve a number of approximations, one of which is
to assume a simple spin-structure for the delta self-energy (achieved
most simply by using the nonrelativistic limit).  
As a prelude to a more complete treatment we consider here the 
vacuum case which may hopefully serve as starting point for the
much more involved in-medium computations.

The organization of the presentation is the following. In section 2 
the calculation of the self-energy and dressed propagator is 
presented. The results are applied in section 3 to computing the
pion-nucleon phase shift in the spin-3/2 isospin-3/2 channel and 
fitting the parameters to observed data. Conclusions and outlook 
are discussed in section 4.

%%%%%%%%%%%%%%%%%%%%%%%%%%%%%%%%%%%%%%%%%%%%%%%%%%%%%%%%%%%%%%%%%%%%%
\section{Calculation of the self-energy and propagator}
%%%%%%%%%%%%%%%%%%%%%%%%%%%%%%%%%%%%%%%%%%%%%%%%%%%%%%%%%%%%%%%%%%%%%
From expression (\ref{pnd}) 
the (unregularized) one-loop delta self-energy has the
following form:
\be
\Sigma^{\mu\nu}(p)=-i\frac{g^2}{M^2}
\int \frac{d^4k}{(2\pi)^4} (k^\mu +a\gamma^\mu
\slash{\!\!\!k})\frac{(\slash{\!\!\!p}+\slash{\!\!\!k}+M)}
{(p+k)^2-M^2+i\epsilon}
(k^\nu+a\slash{\!\!\!k}\gamma^\nu)\frac{1}{k^2-m^2+i\epsilon},
\ee
where $M$ is the mass of the
nucleon and $m$ that of the pion. The above expression is divergent
and for regularization we use a Lorentz-scalar form-factor in the
vertex. The form-factor is used only when calculating the imaginary
part of the self-energy, whose real part is then obtained from a
dispersion relation. In this way the correct analytic properties of
of the self-energy and thus causality are assured.

The self-energy can be written in the following general form, which
contains ten independent second-order tensor forms which can be
constructed from the delta's four-momentum $p$, the gamma matrices
and
the metric tensor:
\bea
\Sigma^{\mu\nu}(p)&=&g^{\mu\nu}(\beta_1+\slp\beta_2)
+p^\mu p^\nu (\beta_3+\slp \beta_4)
+\gamma^\mu \gamma^\nu (\beta_5+\slp \beta_6)\nonumber \\
&&+p^\mu \gamma^\nu \slp \beta_7
+p^\nu \gamma^\mu \slp \beta_8
+p^\mu \gamma^\nu \beta_9 +p^\nu \gamma^\mu \beta_{10}.
\label{sigma}\eea
The functions $\beta_i\equiv\beta_i(p^2)$ are Lorentz-scalars
which depend only on $p^2$.

When calculating imaginary parts of $\beta_i$ we start from the
expression for the imaginary part of a loop diagram \cite{Philold}:
\be
{\rm{Im}}\;\Sigma^{\mu\nu}=2\frac{g^2}{M^2}\int\frac{d^4k}{(2\pi)^4}
\Gamma^\mu \;{\rm{Im}}\,G(p+k)\;{\rm{Im}}\,D(k)
\;\Gamma^\nu \theta(p_0+k_0)\theta(-k_0),
\ee
with $\Gamma^\mu$ standing for the spin structure of the coupling 
(including also the form-factor) and
the imaginary parts of the (free) nucleon and pion propagators given
by
\bea
{\rm{Im}}\,D(k)\;\theta(-k_0)&=&-\frac{\pi}{2\sqrt{m^2+\bm{k}\,^2}}
\,\delta(k_0+\sqrt{m^2+\bm{k}\,^2}),\nonumber\\
{\rm{Im}}\,G(p+k)\;\theta(p_0+k_0)&=&-\frac{\pi}{2\sqrt{M^2+
(\bm{p}+\bm{k})^2}}\,
(\slp+\slk+M)\nonumber\\
&&\times\delta(p_0+k_0-\sqrt{M^2+(\bm{p}+\bm{k})^2}).
\eea
The presence of the two delta-functions allows for complete
evaluation
of the integral.
The imaginary parts of functions $\beta_i(p^2)$ are nonzero only 
if $p^2>(M+m)^2$
and in the calculation it is convenient to use $\bm{p}=0$. 
Taking into account the presence of the form-factor $F(p^2)$ 
(with explicit form specified below), the imaginary parts of
coefficients $\beta_i(p^2)$ all have the following form
\be
{\rm{Im}}\,\beta_i(p^2)=\alpha_i(p^2) 
\frac{g^2 F(p^2)^2}{16\pi M^2p^2}\sqrt{(p^2-M^2+m^2)^2-
4m^2p^2}.
\ee 
Introducing the notation
\bea
k_*^2&\equiv &\frac{1}{4p^2}[(p^2-M^2+m^2)^2-4m^2p^2],\nonumber\\
p_*^2 &\equiv & p^2-M^2+m^2,
\eea
for the functions $\alpha_i(p^2)$ we obtain:
\bea
\alpha_1&=&-\frac{Mk_*^2}{3},\nonumber\\
\alpha_2&=&-\frac{k_*^2}{6p^2}(p^2+M^2-m^2),\nonumber\\
\alpha_3&=&\frac{(1+2a)M}{3p^4}[p_*^4-m^2p^2],\nonumber\\
\alpha_4&=&\frac{1}{12p^6}[p_*^4(p^2+3M^2-3m^2)+2m^2p^2(
p^2-3M^2+3m^2)],\nonumber\\
\alpha_5&=&-\frac{aM}{3}(2k_*^2-3am^2),\nonumber\\
\alpha_6&=&-\frac{2}{3}ak_*^2-\frac{a^2}{2p^2}[p_*^4-
m^2(3p^2-M^2+m^2)],\nonumber\\
\alpha_7 &=&-\alpha_8=-\frac{aM}{3p^4}(p_*^4-m^2p^2),
\nonumber\\
\alpha_9 &=& \frac{1}{6p^2}\left[ k_*^2p_*^2+a(
p_*^4+2m^2p^2-3m^2p_*^2)\right],\nonumber\\
\alpha_{10}&=& \frac{1}{6p^2}\left\{ k_*^2p_*^2+3a(1+2a)
[p_*^4-2m^2p^2-m^2p_*^2]\right\}.
\eea
The real part of $\beta_i(p^2)$ is then calculated numerically
using the dispersion relation
\be
{\rm Re}\,\beta_i(p^2)=\frac{\cal P}{\pi}
\int_{(M+m)^2}^\infty \frac{d\sigma^2 \,{\rm Im}\,\beta_i
(\sigma^2)}{\sigma^2-p^2},
\ee
assuring the correct analytic properties of the self-energy.

For the form-factor we used an exponential form:
\be
F(p^2)=\exp\left[-\frac{p^2-(M+m)^2}{\Lambda^2}\right],
\ee
which provides suppression in the kinematical range of 
interest, since the imaginary part of the self-energy in
the considered model is
nonzero only if $p^2 > (M+m)^2$.

In order to be able to solve the Dyson equation for the
dressed delta propagator
\be
G^{\mu\nu}(p)=G_0^{\mu\nu}+G_0^{\mu\alpha}\Sigma_
{\alpha\beta} G^{\beta\nu},\label{Dyson}
\ee
we have to rewrite the self-energy and the free delta propagator
in terms of the following spin-projection operators \cite{Olaf}
(omitting on the left side the
two Lorentz-vector indices):
\bea
P^{3/2}&=&g^{\mu\nu}-
\frac{2p^\mu p^\nu}{3p^2}
-\frac{\gamma^\mu \gamma^\nu}{3}
+\frac{1}{3p^2}(\gamma^\mu p^\nu-p^\mu \gamma^\nu)\slp,
\nonumber\\
P_{11}^{1/2}&=&\frac{\gamma^\mu \gamma^\nu}{3}
-\frac{p^\mu p^\nu}{3p^2}
-\frac{1}{3p^2}(\gamma^\mu p^\nu-p^\mu \gamma^\nu)\slp,
\nonumber\\
P_{22}^{1/2}&=&\frac{p^\mu p^\nu}{p^2},\nonumber\\
P^{1/2}_{21}&=&\frac{1}{\sqrt{3} p^2}
(-p^\mu p^\nu +\gamma^\mu p^\nu \slp),\nonumber\\
P_{12}^{1/2}&=&\frac{1}{\sqrt{3}p^2}
(p^\mu p^\nu-p^\mu \gamma^\nu \slp).\label{projectors}
\eea
The coefficient of each projector has the form $\slp a(p^2)+b(p^2)$, 
where $a(p^2)$ and $b(p^2)$ are Lorentz scalars.
From Eq.\ (\ref{Dyson}) the dressed propagator can be obtained
through
\be
G^{-1}=G_0^{-1}-\Sigma,\label{Dyson1}
\ee
where from (\ref{freepropagator})
\be
G_0^{-1}=(P^{3/2}-2P_{11}^{1/2})(\slp-M_\Delta)
+\sqrt{3}M_\Delta (P_{12}^{1/2}+P_{21}^{1/2}).
\ee
Using the properties of the projectors (\ref{projectors}) one can
invert expression (\ref{Dyson1}) to obtain the dressed delta 
propagator in the following form:
\be
G(p)=\frac{a_1\slp-b_1}{p^2a_1^2-b_1^2}P^{3/2}
+(a_2\slp+b_2)P_{11}^{1/2}
+(a_3\slp+b_3)P_{22}^{1/2}
+(a_4\slp+b_4)P_{12}^{1/2}
+(a_5\slp+b_5)P_{21}^{1/2}.\label{dressedpropagator}
\ee
The coefficients $a_i$ and $b_i$ are functions of $p^2$ and in 
general have a real and imaginary part. The factor of the
projector $P^{3/2}$ in (\ref{dressedpropagator}) is written
explicitly in terms of $a_1$ and $b_1$ which multiply 
(in the form $a_1 \slp+b_1$) that projector from the right
in the expression for $G^{-1}$. The other eight functions 
$a_i,b_i,\;i=2\dots 5$ are obtained from an algebraic system
of eight equations containing the eight coefficients of the 
other four projectors in the expression for $G^{-1}$.
To perform actual 
calculations with this dressed propagator it is necessary to 
write it in the form of the right side of expression (\ref{sigma}).

As a partial check on the correctness of the calculation,
for every set of parameters used,
we numerically computed the sum-rule for the $g^{\mu\nu}\gamma^0$
term of the delta spectral-function. From expression 
(\ref{freepropagator}) it follows that the free 
propagator satisfies
\be
\int_{-\infty}^{\infty} \rho_0(p_0,\bm p) dp_0=1,
\ee
where $\rho_0$ is the imaginary part (devided by $-\pi$)
of the coefficient of the $\gamma^0 g^{\mu\nu}$ term of the
propagator. Numerical evaluation of the corresponding
sum-rule for the dressed propagator, 
which in view of (\ref{projectors}) and notation 
(\ref{dressedpropagator}) can be written as
\be
-\frac{1}{\pi}\int_{-\infty}^{\infty} 
dp_0 p_0 {\rm{Im}}\,\frac{a_1}{p^2a_1^2-b_1^2}=1,
\ee 
showed it to be satisfied to accuracy better than 1\%.

%%%%%%%%%%%%%%%%%%%%%%%%%%%%%%%%%%%%%%%%%%%%%%%%%%%%%%%%%%%%%%%%%%%%%
\section{Pion-nucleon phase-shift in the (3/2,3/2) channel}
%%%%%%%%%%%%%%%%%%%%%%%%%%%%%%%%%%%%%%%%%%%%%%%%%%%%%%%%%%%%%%%%%%%%%

The pion-nucleon phase-shift in the spin-3/2 isospin-3/2 channel 
(more precisey the P$_{33}$ phase-shift) shows a resonance behavior 
which can be explained very well with a diagram 
(see Fig.\ 1) involving the delta
intermediate state \cite{Ericson}. 
%\pagebreak[4]

\vspace*{8mm}
\epsfxsize=6cm
\centerline{\epsffile{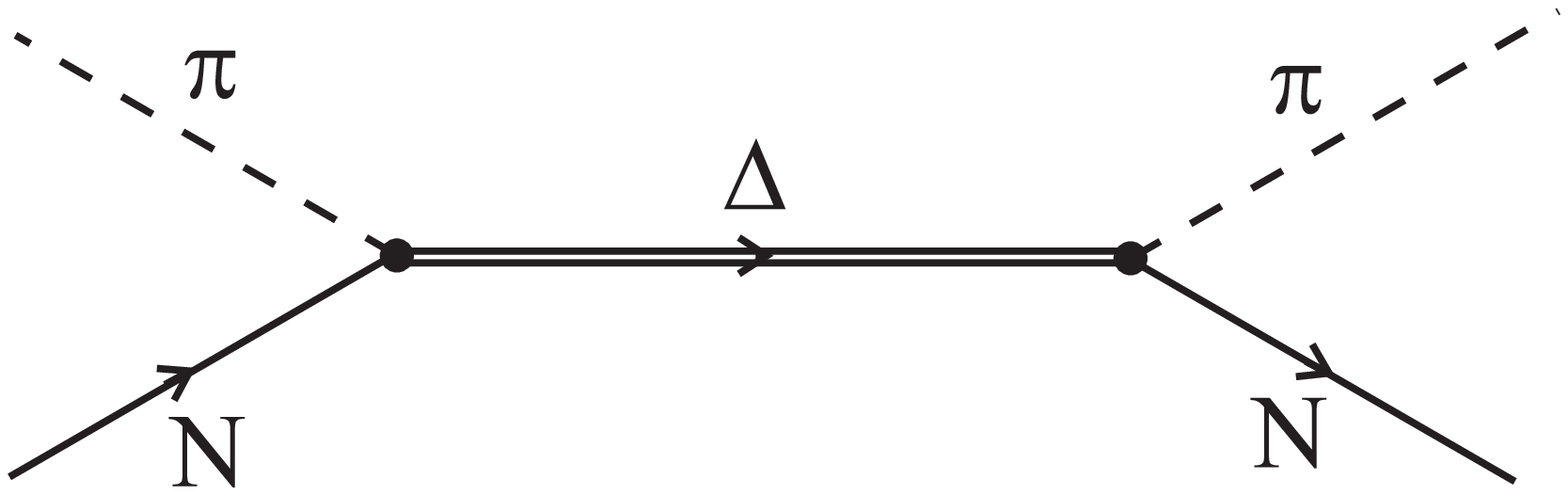}}
\vspace*{3mm}
\begin{center}
Fig.\ 1.\ The pion-nucleon scattering in the spin-3/2 isospin-3/2
channel
can be described well through a delta in intermediate state.
\end{center}

We want to establish whether 
the same good quality fit can be achieved with the present complete
relativistic treatment of the delta. The usual treatment 
\cite{Ericson,Korpa} involves the nonrelativistic limit, while in
Ref.\ \cite{Phil} a relativistic treatment was used, but with an
approximation for the self-energy and dressed delta propagator.

In the nonrelativistic limit the pion-nucleon-delta coupling has
only a p-wave ($\ell =1$) component, while a fully relativistic
vertex includes also a d-wave ($\ell =2$) term. This means that apart
from the P$_{33}$ phase-shift the diagram in Fig.\ 1  
contains the D$_{33}$ phase-shift too.
The general expression for the S-matrix element for scattering
of spin-zero particle by spin-one particle in the center-of-mass
system \cite{Goldberger} is:
\bea
<\hat{e}_f,\frac{1}{2},\nu'|S|\hat{e}_i,\frac{1}{2},\nu>&=&
\sum_{\ell m m' J M}
Y_\ell ^{m'}(\hat{e}_f)
Y_\ell ^{m\,*}(\hat{e}_i)
<\ell,\frac{1}{2};m',\nu'|JM>\nonumber\\
&&\times <\ell,\frac{1}{2};m,\nu|JM> T_\ell^J,\label{smatrix}
\eea
where $\hat{e}_i$ ($\hat{e}_f$) gives the direction of the incoming 
(outgoing) momentum, $\nu$ ($\nu'$) is the spin projection on the
momentum for the incoming (outgoing) state and
\be
T_\ell^J\equiv \frac{1}{2\pi i \rho}\left( 1-e^{2i\delta_\ell^J}
\right).
\ee
$\delta_\ell^J$ is the phase-shift in the considered channel and
$\rho$ is the energy density of states.
Taking $\nu=\nu'=1/2$ and $\hat{e}_i$ to point in the direction of 
the z-axis, only terms with $m=m'=0$ (and $M=1/2$) survive in the 
sum on the right-side of (\ref{smatrix}). Since the delta
intermediate state means that $J=3/2$, the two possible $\ell$
values are 1 and 2. We can eliminate the term involving the
$\ell=2$ phase-shift by choosing a scattering angle such that
$Y_2^0(\hat{e}_f)=0$. This gives for the polar angle $\vartheta=
54.7^{\mbox{\scriptsize o}}$. 

\vspace*{8mm}
\epsfxsize=8cm
\centerline{\epsffile{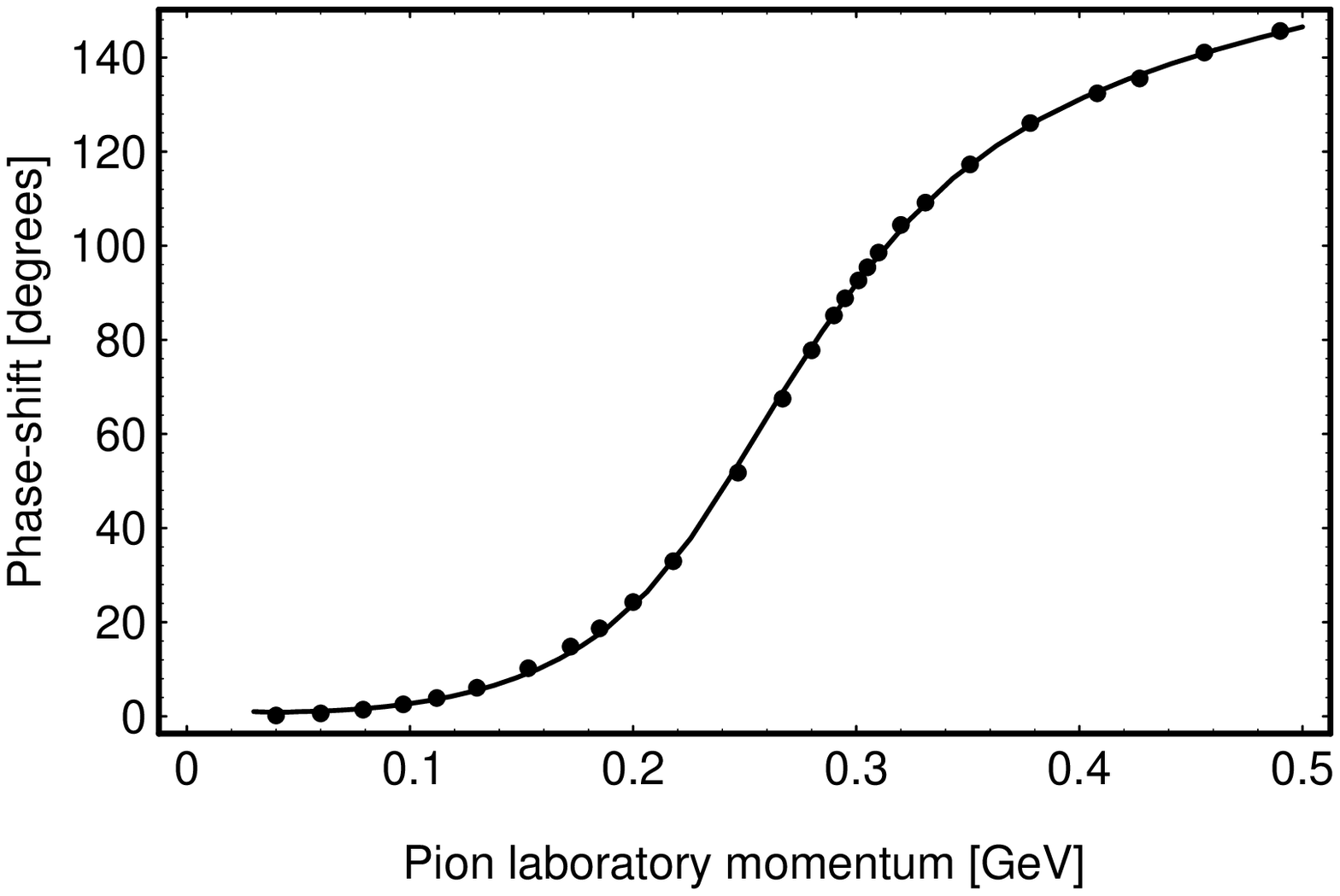}}
\vspace*{3mm}
\begin{center}
Fig.\ 2.\ 
Phase-shift in the spin-3/2 isospin-3/2 channel. The points
represent measurement results from Ref.\ \protect{\cite{Koch}},
while the line is our calulated result with values of
parameters given in text.
\end{center}

The scattering amplitude of the diagram in Fig.\ 1 was calculated
numerically, using the dressed delta propagator and the 
pion-nucleon-delta vertex corresponding to interaction (\ref{pnd}).
The P$_{33}$ pion-nucleon phase-shift was reproduced with excellent
agreement (shown in Fig.\ 2) up to pion laboratory momentum of 
$500\;$MeV. The parameters used were the following: coupling 
$g=19$, cut-off $\Lambda=0.97\;$GeV, bare
mass of the delta $M_\Delta=1.279\;$GeV. 
Based on the considered P$_{33}$ phase-shift it is not possible 
to fit uniquely the value of the off-shell 
parameter $a$ in the interaction (\ref{pnd}). Values of $a$  
in the range from $-1$ to $0$ give excellent fits, but 
much larger or much smaller values completely destroy the
agreement.
The cut-off $\Lambda$ is seemingly much larger than the one 
obtained in Ref.\ \cite{Korpa}, where a non-relativistic approach 
was used. A direct comparison of the two values is, however, not
possible because of the different arguments used 
(in Ref.\ \cite{Korpa} it was the pion's three-momentum which 
appeared in the form-factor). An approximate correspondence of 
the two functional forms is established for 
$\Lambda_{\protect{\cite{Korpa}}}=\Lambda/\sqrt{2}=0.69\;$GeV, still
a considerably larger value than obtained in Ref.\ \cite{Korpa}.
The form-factor in the present approach, nevertheless, is rather
soft, similarly to the result of Ref.\ \cite{Phil}. 
We remark that a very good fit could be obtained by using a somewhat
larger value of the coupling $g=20$ with a slightly softer
form-factor
with $\Lambda=0.92\;$GeV, and a bare mass of the delta $M_\Delta=
1.273\;$GeV. These values of $g$ and $\Lambda$ give a similar value
for the on-mass-shell coupling as the values $g=19$ and $\Lambda=
0.97\;$GeV.

%%%%%%%%%%%%%%%%%%%%%%%%%%%%%%%%%%%%%%%%%%%%%%%%%%%%%%%%%%%%%%%%%%%%%
\section{Conclusions}
%%%%%%%%%%%%%%%%%%%%%%%%%%%%%%%%%%%%%%%%%%%%%%%%%%%%%%%%%%%%%%%%%%%%% 
Although the full tensor structure of the pion-nucleon-loop delta
self-energy in the relativistic treatment is rather involved, it 
can be calculated in a straightforward way. It is convenient to 
use the dispersion relation satisfied by the self-energy and 
calculate first its imaginary part (more precisely the imaginary
parts of the Lorentz-scalar coefficients determining it). These 
were calculated in an analytic form, using the most general 
pion-nucleon-delta coupling, involving also the off-shell 
parameter. Using an exponential form-factor to regularize the
self-energy the real parts of the ten Lorentz-scalar functions
were calculated numerically, as well as the Lorentz-scalar
coefficients of the dressed propagator, obtained from solving the
Schwinger-Dyson equation. A sum-rule for the dressed delta 
propagator was checked numerically.

The dressed delta propagator was used to calculate the pion-nucleon
P$_{33}$ phase-shift in the spin-3/2 isospin-3/2 channel. An
excellent fit was obtained using a rather soft pion-nucleon-delta
form-factor and a coupling whose on-shell value is close to the one
determined from the decay width of the delta. The off-shell parameter
$a$ could not be determined uniquely from a fit to this process,
since
values between $-1$ and $0$ give practically the same result (much
larger or smaller values destroy the fit). 
It would be of interest to
examine how observables in some other processes involving the
delta (e.g.\ Compton scattering on a nucleon, pion electroproduction, 
etc.) are affected by the off-shell parameter and in general by the
present more complete treatment of the delta propagator.

\acknowledgments

We thank Alex Korchin for useful discussions and Olaf Scholten for 
providing information about his work. This research was supported 
in part by the Hungarian Research Foundation (OTKA) grant T16594.

\end{document}